\documentclass[aps,prl,twocolumn,superscriptaddress]{revtex4}

\usepackage[T1]{fontenc}
\usepackage[latin9]{inputenc}
\usepackage{amsmath}
\usepackage{amssymb}
\usepackage{graphicx}
\usepackage{hyperref}
\usepackage{natbib}
\usepackage{esint}
\usepackage{times}
\usepackage{helvet}
\usepackage{braket}
\usepackage{bbm}
 
\makeatletter
\begin{document}

\title{
Cluster State Generation with Spin-Orbit Coupled Fermionic Atoms in Optical Lattices
}

\author{M. Mamaev}
\affiliation{JILA, NIST and Department of Physics, University of Colorado, Boulder, USA}
\affiliation{Center for Theory of Quantum Matter, University of Colorado, Boulder, CO 80309, USA}
\author{R. Blatt}
\affiliation{Institute for Quantum Optics and Quantum Information, Austrian Academy of Sciences,
Technikerstra\ss e 21a, 6020 Innsbruck, Austria}
\affiliation{Institut f\"{u}r Experimentalphysik, Universit\"{a}t Innsbruck, Technikerstra\ss e 25, 6020 Innsbruck, Austria}
\author{J. Ye}
\affiliation{JILA, NIST and Department of Physics, University of Colorado, Boulder, USA}
\author{A. M. Rey}
\affiliation{JILA, NIST and Department of Physics, University of Colorado, Boulder, USA}
\affiliation{Center for Theory of Quantum Matter, University of Colorado, Boulder, CO 80309, USA}

\begin{abstract}{Measurement-based quantum computation, an alternative paradigm for quantum information processing, uses simple measurements on qubits prepared in cluster states, a class of multiparty entangled states with useful properties. Here we propose and analyze a scheme that takes advantage of the interplay between spin-orbit coupling and superexchange interactions, in the presence of a coherent drive, to deterministically generate macroscopic arrays of cluster states in fermionic alkaline earth atoms trapped in three dimensional (3D) optical lattices. The scheme dynamically generates cluster states without the need of engineered transport, and is robust in the presence of holes, a typical imperfection in cold atom Mott insulators. The protocol is of particular relevance for the new generation of 3D optical lattice clocks with coherence times $> 10$ s, two orders of magnitude larger than the cluster state generation time. We propose the use of collective measurements and time-reversal of the Hamiltonian to benchmark the underlying Ising model dynamics and the generated many-body correlations.}\end{abstract}

\maketitle

Entanglement, the characteristic trait of quantum mechanics, is a vital resource for quantum information processing~\cite{nielsen2002quantumInfoSeminal}, quantum communications~\cite{gisin2002quantumCommSeminal} and enhanced metrology~\cite{roos2006quantumMetSeminal}. These applications often require multipartite entangled states, which can be difficult to create and intrinsically fragile to noise and decoherence. Nevertheless, there exists a special class of multipartite entangled states called cluster states, which can be robust to adverse effects on a subset of their logical qubits~\cite{raussendorf2001clusterSeminal,raussendorf2003clusterSeminal,briegel2009clusterSeminal}. This intrinsic robustness, and the state entanglement properties, make cluster states in two (or three) dimensions a resource for one-way quantum computing, where a computation is realized by a sequence of single-qubit measurements on the initial cluster state. Besides their appeal in quantum computation, cluster states have been a playground for the study of many-body and statistical physics~\cite{briegel2009clusterSeminal}, graph theory~\cite{hein2004clusterGraph}, topological codes~\cite{raussendorf2007clusterTopo}, and mathematical logic~\cite{van2008clusterMathLogic}.

Cluster state generation has been reported in proof-of-principle experiments using frequency down-conversion techniques~\cite{chen2014clusterDownConversion,lu2007clusterFrequency,walther2005clusterFrequency}, photonic qubits~\cite{chen2007clusterPhotonic,kiesel2005clusterPhotonQubit}, continuous-variable modes of squeezed light~\cite{yokoyama2013clusterCV,yukawa2008clusterCVexpt}, semiconductor quantum dots~\cite{schwartz2016clusterCavityQED} and trapped ions~\cite{lanyon2013clusterTrappedIons}. In addition, coherent entangling-disentangling
evolution via controlled collisions was reported in cold atom Mott insulators~\cite{mandel2003clusterOpticalLattice}, an experiment that stimulated theoretical work towards cluster state generation~\cite{daley2008cluster,vaucher2008clusterOpticalLattice,kuznetsova2012cluster,treutlein2006clusterOpticalLattice,zwierz2009clusterDipoleBlockadeAtomic}. However, a scalable, deterministic source of cluster states needs yet to be realized.

Here we propose a scheme for preparing macroscopic cluster state arrays ($\sim 10^3$ qubits ) in one, two and three dimensions. Our protocol uses a combination of superexchange and spin-orbit coupling to engineer nearest-neighbor Ising interactions. In this implementation cluster states naturally emerge during time evolution without the need of controlled collisions in spin dependent lattices~\cite{daley2008cluster}, while maintaining robustness to imperfect filling. While full tomography is not yet feasible in macroscopic systems, we propose the use of  many-body echoes to  probe the cluster state quality.
While our protocol is general and applicable to ultracold atomic systems interacting via contact~\cite{Gross2017} or engineered interactions (e.g. via an optical cavity)~\cite{landig2016cavity,camacho2017cavity}, it is particularly relevant for current 3D atomic lattice clocks~\cite{campbell2017clockSeminal,marti2018clock,goban2018lifetimeLimitation} operated with 
fermionic alkaline earth atoms (AE). These atoms offer untapped opportunities for precision metrology~\cite{ludlow2015alkaliearth} and quantum information ~\cite{Daley2011,Cazalilla2014,goban2018lifetimeLimitation}, since they possess a unique atomic structure featuring an ultra-narrow clock transition with $>100$s lifetimes, and a fully controllable, magnetic field insensitive hyperfine manifold. The demonstrated capability to generate spin-orbit coupling (SOC) in AEs \cite{wall2016soc,kolkowitz2017soc,bromley2018soc,livi2016soc,galitski2013soc}, together with near-term experimentally accessible single-site addressability and control of SOC via accordion lattices~\cite{al2010accordion,williams2008accordion,yeAccordionRef}, may enable the first realization of a large-scale one-way quantum computer in ultracold atoms using our protocol.

\textit{Model.}
\begin{figure}
\centering
\includegraphics[width=1\linewidth]{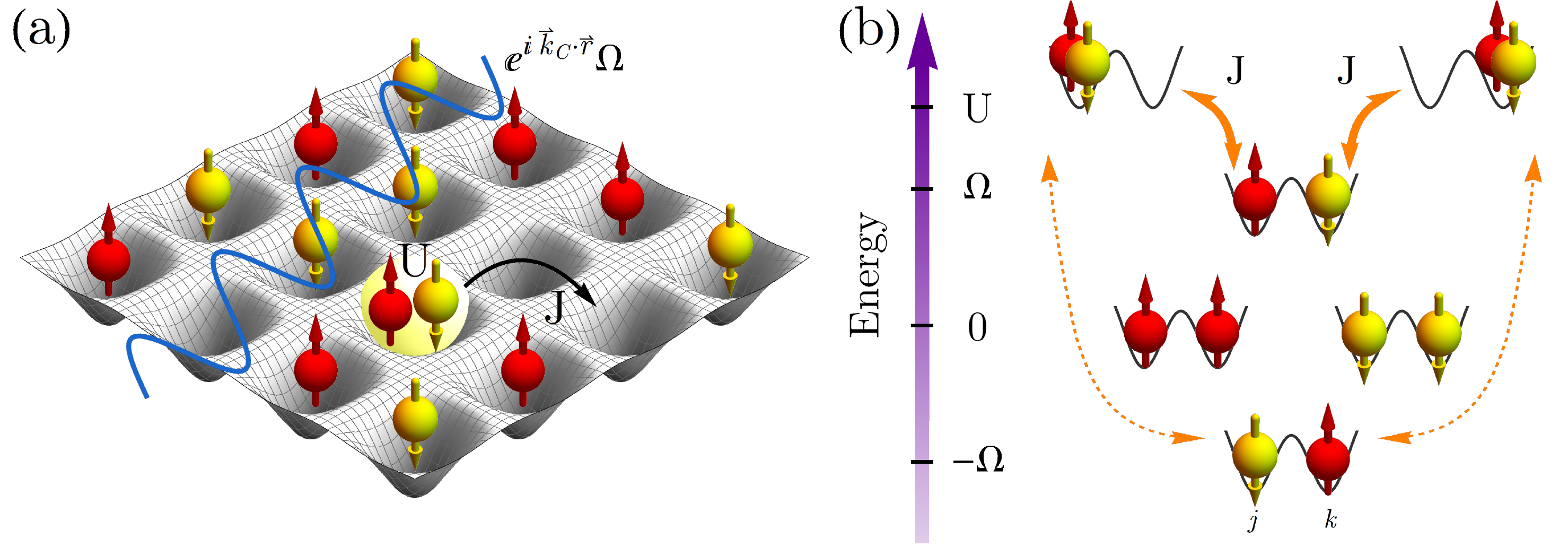}
\caption{\textbf{(a)} Schematic of Fermi-Hubbard dynamics in an optical lattice, characterized by nearest-neighbour tunneling energy $J$, and onsite repulsion $U$. A resonant laser with Rabi frequency $\Omega e^{i \vec{k}_{\mathrm{C}}\cdot \vec{r}}$ interrogates the internal levels while transferring momentum to the atoms(SOC). \textbf{(b)} Superexchange mechanism. The lowest(highest)-energy single-particle levels are in a staggered spin configuration due to the SOC. One type of virtual tunneling is suppressed by an energy cost $U+\Omega$, while another is near-resonant with cost $U-\Omega$ (for $U \simeq \Omega$). Zero-energy states play no role due to Pauli exclusion prohibiting tunneling.}
\label{fig_Schematic}
\end{figure}
Consider $N$ neutral fermionic atoms prepared in two long-lived internal states, denoted by $g, e$ (e.g. optical clock states or hyperfine nuclear spin states), trapped in a deep cubic optical lattice of $L$ sites. We operate in the ultracold regime where  only the lowest Bloch band is populated. The internal levels are continuously driven by a resonant laser (via optical or Raman transitions) with wavevector $\vec{k}_{\mathrm{C}}$ and Rabi frequency $\Omega e^{i \vec{k}_{\mathrm{C}}\cdot \vec{r}}$ at lattice position $\vec{r}$. The drive imprints a site-dependent phase $\phi_j=\vec{k}_{\mathrm{C}}\cdot \vec{r}_{j}$, which transfers momentum to the atoms and generates spin-orbit coupling~\cite{wall2016soc}. Here, $\vec{r}_{j}=(m,n,l) a$, with $a$ the lattice spacing and $m,n,l$ integers.
By going to a dressed basis $\sigma\in \{\uparrow,\downarrow\}$ defined by the rotated states $\ket{\uparrow}=(\ket{e}-i\ket{g})/\sqrt{2} $ and $\ket{\downarrow}=(\ket{e}+i\ket{g})/\sqrt{2}$, (c.f. Supplementary~\ref{app_FermiHubbardRotated}), the Hamiltonian is described by the following Fermi-Hubbard model ($\hbar=1$):
\begin{equation}
\begin{aligned}
\label{eq_FermiHubbard}
    \hat{H}=&-J \sum_{\langle j,k\rangle,\sigma}\left(\hat{c}_{j\sigma}^{\dagger}\hat{c}_{k\sigma}+h.c.\right)+U\sum_{j}\hat{n}_{j\uparrow}\hat{n}_{j\downarrow}\\
    &+\frac{\Omega}{2}\sum_{j}e^{i\pi j}\left(\hat{n}_{j\uparrow}-\hat{n}_{j\downarrow}\right).
\end{aligned}
\end{equation}
Here, $\hat{c}_{j\sigma}$ annihilates an atom of spin $\sigma$ on site $j$, $\hat{n}_{j\sigma}=\hat{c}_{j\sigma}^{\dagger}\hat{c}_{j\sigma}$, and $\langle j,k \rangle$ indexes nearest-neighbours (no double-counting). $J$ is the hopping amplitude between sites, and $U>0$ is the Hubbard repulsion [See Fig.~\ref{fig_Schematic}(a)]. We have assumed that the phase between neighbouring sites is  $\phi_{j}-\phi_{k} =\pi$, corresponding to $a \vec{k}_{\mathrm{C}}\cdot \hat{\alpha}=\pi$ for every unit vector $\hat{\alpha}\in \{\hat{x},\hat{y},\hat{z}\}$ along which tunneling is permitted, so that the sign of the Rabi drive alternates between them. This corresponds to inducing an effective gauge field with relative flux $\pi$, if one visualizes the spin as an additional synthetic dimension~\cite{celi2014syntheticGauge}.

At half filling $N=L$, $\Omega/J \gg 1$ and $U/J \gg 1$, a Mott insulator is formed with suppression of doubly-occupied sites. The strong drive favours staggered spin order, and competes/cooperates with virtual second-order tunneling processes (superexchange). Instead of typical antiferromagnetic interactions~\cite{anderson1950superexchange,trotzky2008superexchange}, SOC transforms the Hamiltonian into a dominant nearest-neighbour Ising interaction. The underlying mechanism is depicted in Fig.~\ref{fig_Schematic}(b). If two neighbouring sites are in the single-particle ground-state $\ket{\downarrow\uparrow}$, double occupancy after a tunneling event will cost an energy penalty  $+\Omega$ due to the alternating Rabi drive, and an additional penalty $+U$ due to Hubbard repulsion, creating a large energy gap $\Omega+U$. If the particles are instead in the excited state $\ket{\uparrow\downarrow}$, virtual tunneling costs $-\Omega+U$ which can be made near-resonant for $\Omega \simeq U$. The states $\ket{\uparrow\uparrow}$ and $\ket{\downarrow\downarrow}$ cannot tunnel due to Pauli exclusion. The effective superexchange Hamiltonian for our system (c.f. Supplementary~\ref{app_Superexchange}) becomes $\hat{H}_{\mathrm{se}}= \hat{H}_{\mathrm{se}}^{(1)}+\hat{H}_{\mathrm{se}}^{(2)}$ with:
\begin{equation}
\label{eq_SuperexchangeHamiltonian}
    \hat{H}_{\mathrm{se}}^{(1)}\equiv\frac{4J^2 U}{\Omega^2-U^2}\sum_{\langle j,k\rangle}\hat{S}_{j}^{z}\hat{S}_{k}^{z}+\left(\Omega+D\frac{4J^2 \Omega}{\Omega^2-U^2}\right)\sum_{j}\hat{S}_{j}^{z},\end{equation}
where $\hat{S}_{j}^{\alpha}$ are spin-1/2 operators, and $D$ is the dimensionality (i.e. $D=2$ for 2D tunneling). There is an additional interaction $\hat{H}_{\mathrm{se}}^{(2)}\equiv\frac{2J^2}{U}\sum_{\langle j,k \rangle}(\hat{S}_{j}^{+}\hat{S}_{k}^{+}+h.c.)$, but its contribution to unitary evolution is rendered negligible in our parameter regime by the SOC, which forces the states affected by $\hat{H}_{\mathrm{se}}^{(2)}$ to pick up a high-frequency phase $\sim e^{-2 i\Omega t}$ from the drive, making their off-diagonal terms in the unitary proportional to $\sim J^2/(\Omega U)$ and thus negligible (c.f. Supplementary~\ref{app_SpinEcho}). The superexchange mapping is exact in the limit of $U/J\to \infty$, and $|\Omega-U|/J \to \infty$ to avoid higher-order processes (see Supplementary~\ref{app_ModelAgreement} for benchmarking).

\textit{Cluster states.}
\begin{figure*}[t]
\centering
\includegraphics[width=0.8\linewidth]{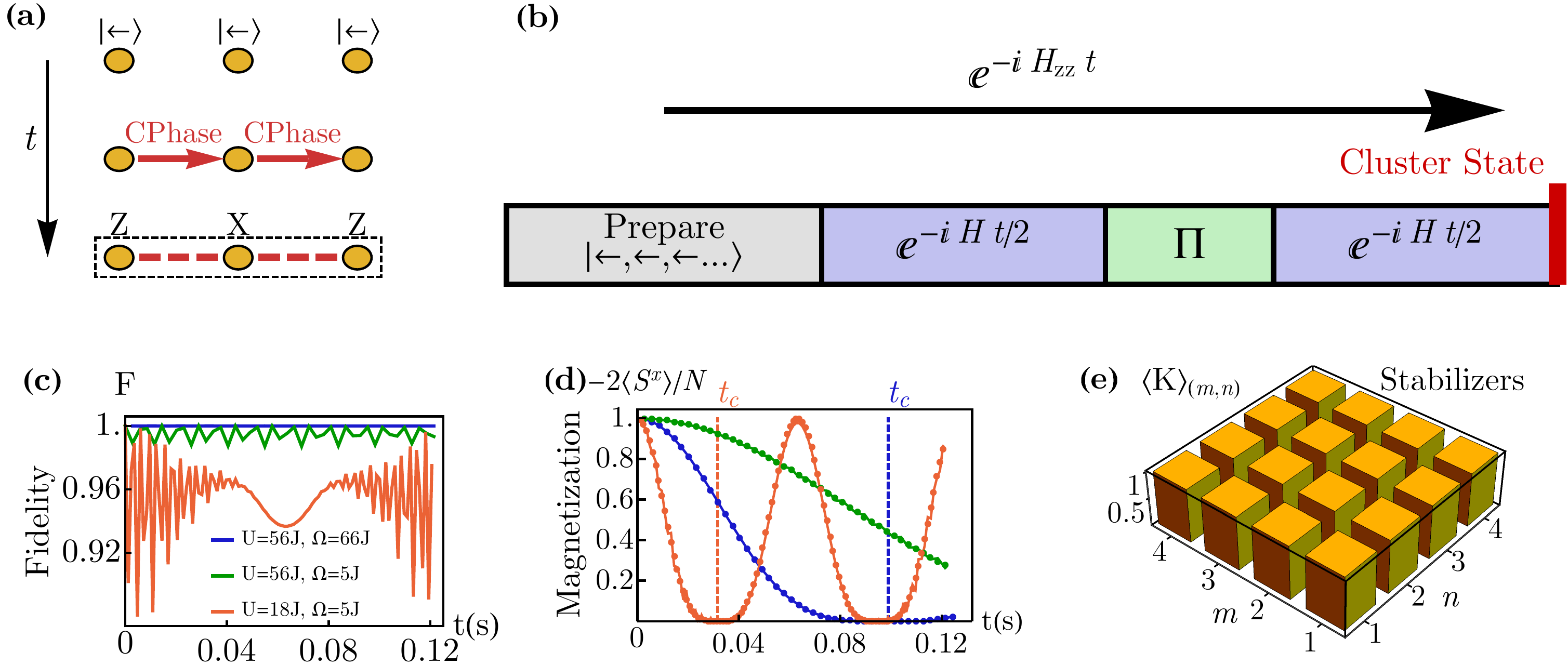}
\caption{ \textbf{(a)} Cluster state schematic. Controlled phase gates are applied to nearest-neighbour pairs. The resulting correlations are local stabilizer operators (here in 1D). \textbf{(b)} Protocol for generating cluster states. The system is evolved with a spin-echo to simulate an Ising interaction. \textbf{(c)} Fidelity $F=|\langle \psi(t)|\psi(t)\rangle_{\mathrm{zz}}|^2$ between an Ising Hamiltonian evolution and our protocol from Eq.~\eqref{eq_IsingEvolution}, using the superexchange model. Parameters are $J/(2\pi) = 28$ Hz for $U/J = 56$, and $J/(2\pi)=66$ Hz for $U/J = 18$. System size is $L=4\times 4$ with 2D tunneling. \textbf{(d)} Time-evolution of $\langle\hat{S}^{x}\rangle$ for our protocol (lines) and ideal Ising model (dots). Cluster times $t_{\mathrm{c}}=\pi/J_{\mathrm{zz}}$ are indicated in matching color. Note that $t_c$ is shorter for the red line because it has a higher $J$, and thus higher $J_{\mathrm{zz}}\sim J^2$. \textbf{(e)} 2D cluster correlations at half filling with $L=4\times 4$, $U/J = 56$, $\Omega/J=66$.}
\label{fig_Cluster}
\end{figure*}
A cluster state $\ket{\psi_{c}}$ is a many-body quantum resource state, characterized by localizeable entanglement. It can be generated by applying a controlled phase gate on every pair of neighbouring sites $\langle j,k\rangle$: $\exp[-i(\hat{S}_{j}^{z}\hat{S}_{k}^{z} + \frac{1}{2}\hat{S}_{j}^{z} + \frac{1}{2}\hat{S}_{k}^{z})\pi]\ket{\leftarrow}_{j}\ket{\leftarrow}_{k}$ where $\ket{\leftarrow}_{j}=(\ket{\uparrow}_{j}-\ket{\downarrow}_{j})/\sqrt{2}$ [See  Fig.~\ref{fig_Cluster}(a)]. Logic gates can be implemented by consecutive measurements on the cluster state, permitting a platform for quantum computation that needs no entanglement generation besides the initial state. A 2D cluster state is sufficient for universal computation~\cite{briegel2009clusterSeminal}, while a 3D state also has significant fault-tolerance on the order of 25\% error~\cite{barrett2010cluster3DErrors}.
We propose to use the Ising interaction in Eq.~\eqref{eq_SuperexchangeHamiltonian} applied to an initial state $\ket{\psi(0)}=\ket{\leftarrow,\leftarrow,\cdots\leftarrow}$ to realize a cluster state. We bring the drive close to resonance, $\Omega \simeq U$, making the $\hat{S}^{z}_{j}\hat{S}^{z}_{j+1}$ term large enough to access cluster states on experimentally viable timescales, as discussed in the last part of this Letter. The single-particle terms in the Hamiltonian can be removed with a spin-echo: We evolve to a halfway time, make a $\pi$-pulse $\hat{\Pi}=e^{-i \pi \hat{S}^{x}}$ and evolve for the second half, undoing any on-site rotations [See Fig.~\ref{fig_Cluster}(b)]:
\begin{equation}
\begin{aligned}
\label{eq_IsingEvolution}
    \ket{\psi(t)}_{\mathrm{zz}}&=e^{-i \hat{H}_{\mathrm{se}}t/2}\hat{\Pi}e^{-i \hat{H}_{\mathrm{se}}t/2}\ket{\psi(0)}\approx e^{-i \hat{H}_{\mathrm{zz}}t}\ket{\psi(0)},\\
    \hat{H}_{\mathrm{zz}}&=J_{zz}\sum_{\langle j,k\rangle}\hat{S}_{j}^{z}\hat{S}_{k}^{z},\>\>\>\>J_{zz}=\frac{4J^2 U}{\Omega^2-U^2}.
\end{aligned}
\end{equation}
Evolving under the Ising interaction to the cluster time, $t_{\mathrm{c}} = \pi/J_{\mathrm{zz}}$ implements the controlled phase gates needed.

At half filling, the protocol prepares an almost perfect cluster state (up to single-particle rotations) for appropriate parameters. Figs.~\ref{fig_Cluster}(c),(d) compare the protocol to the ideal Ising model with fidelity and collective $\hat{S}^{x}=\sum_{j}\hat{S}_{j}^{x}$ observables. 

A cluster state $\ket{\psi_{c}}$ can be equivalently defined as an eigenstate of stabilizer operators~\cite{raussendorf2003clusterSeminal}. These are local multi-body operators that quantify the localizeable entanglement in the state,
\begin{equation}
\label{eq_ClusterStabilizerDefinition}
    \langle \text{K}\rangle_{j}\equiv 2^{2D+1}\bra{\psi}\hat{S}_{j}^{x}\prod_{\langle j,k\rangle}\hat{S}_{k}^{z}\ket{\psi}=1 \text{ for }\ket{\psi}=\ket{\psi_{\mathrm{c}}}.
\end{equation}
The closeness of these stabilizer expectation values, which we call cluster correlations hereafter, to $\langle K\rangle_{j}=1$ in a given region of the lattice is a metric of the cluster state quality there~\cite{raussendorf2003clusterSeminal}. There is no significant distinction between $\langle K\rangle_{j} = \pm 1$, since the two can be interchanged with an $\hat{S}^{z}$ rotation, and we take absolute values when all stabilizers are negative. In Fig.~\ref{fig_Cluster}(e), we show stabilizers for the superexchange model in 2D at half filling. This acts as a better metric than global state fidelity, because the localized nature of cluster state entanglement still permits computation using a region of the lattice if some other, unconnected region is corrupted.

\textit{Imperfect Mott insulator.}
\begin{figure}
\centering
\includegraphics[width=1\linewidth]{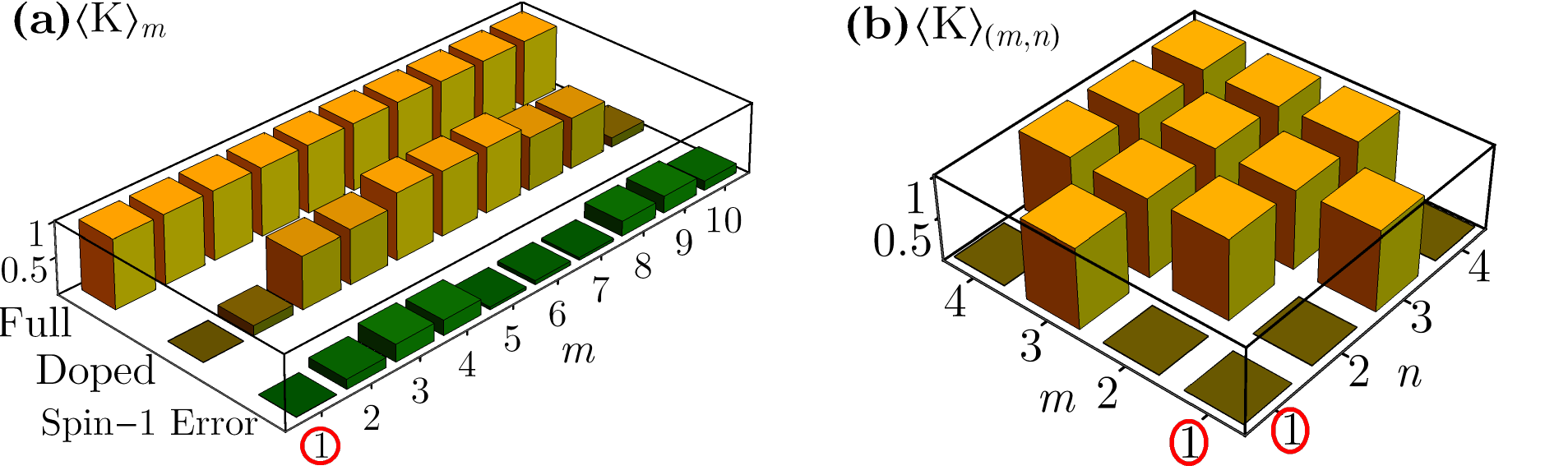}
\caption{\textbf{(a)} Cluster correlations for an $L=10$ system with 1D tunneling at half filling (Full), and with one vacancy initially on site $m=1$ (Doped). Orange plots are computed with Fermi-Hubbard. The green plots quantify how  we would have overestimated the correlations if we had instead used an approximate spin-1 model (c.f. Supplementary~\ref{app_Spin1}). Parameters are $J/(2\pi)=22$ Hz, $U/J=115$, $\Omega/J=140$. \textbf{(b)} Cluster correlations for a 2D system $L=4\times 4$ with one vacancy at $(m,n)=(1,1)$. The spin-1 is used due to the numerical complexity of the Fermi-Hubbard. While it overestimates the correlations, qualitatively the hole remains localized. Parameters are $J/(2\pi)=28$ Hz, $U/J=56$, $\Omega/J=66$.}
\label{fig_Holes}
\end{figure}
A major source of error in experiments is the presence of vacancies in the initial state, which can move and disrupt the correlations. In our implementation they are kept localized by the staggered energy structure imposed by the drive [c.f. Fig.~\ref{fig_Schematic}(b)]. Tunneling into an adjacent empty site costs $\pm\Omega$, and is thus inhibited. While an empty site still destroys the entanglement with its neighbours, other non-adjacent sites can maintain cluster correlations.

Fig.~\ref{fig_Holes}(a) compares cluster correlations for a half-filling sample and a doped array for a 1D system (tunneling allowed along one direction). Sites away from the hole maintain high stabilizer values. A similar result is seen in Fig.~\ref{fig_Holes}(b) for 2D. Given the complexity of solving full Fermi-Hubbard dynamics in this case, we instead use an effective spin-1 model to account for holes (c.f. Supplementary~\ref{app_Spin1}). While that model overestimates the correlations at sites affected by the vacancies [see green plot in Fig.~\ref{fig_Holes}(a)], overall it shows that away from them the correlations persist.

In addition to the above benchmarks, we also compute robustness of stabilizers to increasing system size and external confinement (Supplementary~\ref{app_SizeAndConfinement}).

\textit{Collective cluster measurements and OTOCs.}
Probing stabilizers directly requires measurements of multi-body correlations with single-site resolution. While the resolution is required for one-way quantum information processing, at least for initial test-bed experiments, it is possible to partially bypass this requirement by using inherent properties in the Ising model combined with global probes. Notice that,
\begin{align}
    &\langle \text{K}\rangle_{j}(t)= 2^{2D+1}\bra{\psi(0)}e^{i \hat{H}_{zz}t}\left(\hat{S}_{j}^{x}\prod_{\langle j,k\rangle}\hat{S}_{k}^{z}\right)e^{-i\hat{H}_{zz}t}\ket{\psi(0)},\notag \\
    &= 2(-1)^{D} \bra{\psi(0)}e^{i \hat{H}_{zz}t}\left(e^{-i\hat{H}_{zz}t_{c}}\hat{S}_j^{x}e^{i \hat{H}_{zz}t_{\mathrm{c}}}\right)e^{-i\hat{H}_{zz}t}\ket{\psi(0)},
\label{eq_ClusterCollectiveProtocol}
\end{align}
implying that the many-body measurement can be replaced with a local one by evolving to twice the cluster time instead (c.f. Supplementary~\ref{app_IsingLocalMeasurements}). Measuring over a region $\hat{S}_R^{x}=\sum_{j\in R}\hat{S}_{j}^{x}$ yields mean values of cluster correlations in $R$, which offers a metric for cluster state quality there. This does not contain information about the entire state, but is sufficient to gauge fidelity of computation using the region $R$. While the sign of the Ising interaction inside the brackets of Eq.~\eqref{eq_ClusterCollectiveProtocol} does not matter, the time-reversal of the Hamiltonian can be implemented, thanks to the tunability of the interaction, providing additional benchmarking capability and a more objective comparison.

\begin{figure}
\centering
\includegraphics[width=0.65\linewidth]{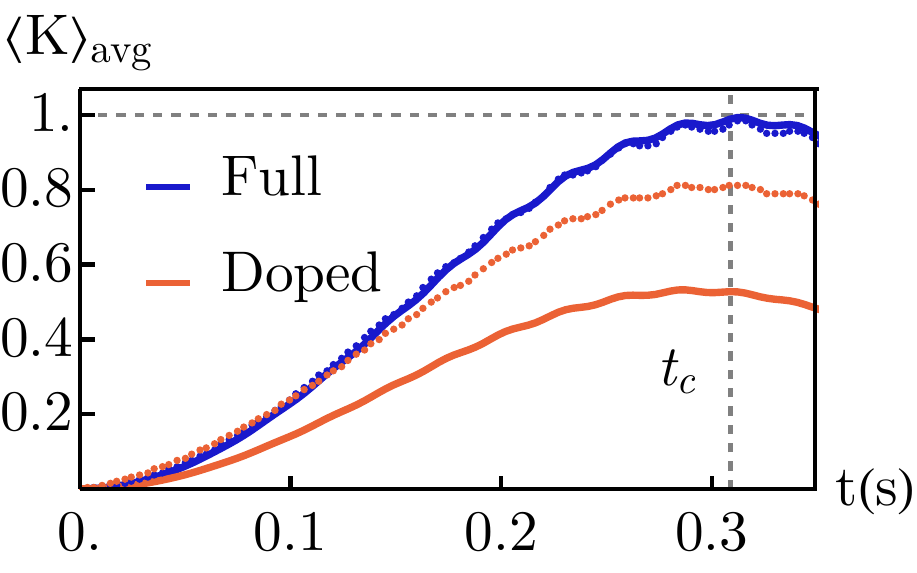}
\caption{Dynamics of average cluster correlations for a 1D tunneling lattice of $L=8$, at half-filling (blue) and one vacancy on site $j=1$ (red), using Fermi-Hubbard. Solid lines generate the cluster state directly, and measure the local correlators. We then average across all sites. Dotted lines use the time-reversal protocol of Eq.~\eqref{eq_ClusterCollectiveProtocol}, relying upon only collective measurements of $\langle\hat{S}^{x}\rangle$. Parameters are $J/(2\pi)=22$ Hz, $U/J=115$, $\Omega/J=140$.}
\label{fig_ClusterCollective}
\end{figure}

After evolving to the cluster time, we quench the drive, $\Omega\to\sqrt{2U^2-\Omega^2}$, causing the interaction to flip its sign, $\hat{H}_{\mathrm{zz}}\to-\hat{H}_{\mathrm{zz}}$ [the Ising model is realized with the spin-echo of Eq.~\eqref{eq_IsingEvolution}]. If the mapping between the Fermi-Hubbard and superexchange were exact, then at $t = t_{\mathrm{c}}$ we implement a unitary reversal and measure ideal cluster correlations. Doping or non-ideal implementation of the Ising would yield lower values. Fig.~\ref{fig_ClusterCollective} compares the dynamics of cluster correlations with exact many-body measurements and the collective measurement $\langle\hat{S}^{x}\rangle(t)$ ($R=N$). With half-filling, we see near-perfect agreement. For a doped array, the collective measurements overestimate the correlations, but still maintain the overall trend.

The goal of the above protocol is to gauge the cluster state quality. To actually implement quantum computation after cluster states are generated requires a non-trivial measurement sequence~\cite{raussendorf2001computational,danos2007measurement}. Since the focus of this work is cluster state generation instead of one-way quantum computation, we leave the details of the latter to future work.

As a side remark, the ability to generate time-reversed evolution allows to measure out-of-time-ordered correlations (OTOCs)~\cite{shenker2014otoc,swingle2016otocs,bollinger2017otocs}. An OTOC is defined as $C_{WV}(t)=\langle\hat{W}(t)^{\dagger}\hat{V}^{\dagger}\hat{W}(t) \hat{V}\rangle$, where $\hat{W}, \hat{V}$ are commuting operators and  $\hat{W}(t)=e^{i \hat{H}t} \hat{W} e^{-i \hat{H} t}$. OTOCs quantify how quantum information is scrambled over many-body degrees of freedom after a quench~\cite{shenker2014otoc}. OTOCs have also been considered a proxy of quantum chaos~\cite{maldacena2016otocChaos}. In our system, OTOCs can be measured if we choose $\hat V=\hat{S}^{x}$, since 
$\langle \hat{W}(t)^{\dagger}\hat{V}^{\dagger}\hat{W}(t) \hat{V}\rangle=-L\langle \hat{W}(t)^{\dagger}\hat{V}^{\dagger}\hat{W}(t) \rangle/2$, and $\hat{W} = e^{-i \theta \hat{S}^{x}}$ is a collective rotation for some angle $\theta$, which is straightforward to realize experimentally. Different OTOCs can be measured by using different rotation axes or angles.

\textit{Experimental parameters and implementations.}
One of the most promising systems to implement our proposal is the 3D optical lattice clock, operated with fermionic Strontium-87 atoms in a cubic lattice at the 'magic-wavelength' $a\approx 406$ nm~\cite{ye2008clockMagic}. Along the directions where we want tunneling, we assume lattice confinement of $V_0/E_{r} \approx 15-20$ ($E_r$ the recoil energy), to obtain $J \sim 10\times 2\pi$ Hz, and much deeper confinement $V_0/E_{r}\gtrsim 100$ along other directions. For a scattering length $a_{eg}^{-}\approx 69 a_0$~\cite{goban2018lifetimeLimitation} ($a_0$ the Bohr radius), interaction strength is $U/J \sim 100$. The Rabi frequency needs to satisfy both $\Omega \simeq U$, and $|U-\Omega|/J \gg 1$ to allow fast cluster state generation $t_{\mathrm{c}}\sim 0.1$ s compared to the current experimental coherence time of $\sim 10$ s~\cite{goban2018lifetimeLimitation}, and to guarantee the validity of the superexchange model.

The spin degree of freedom one can be encoded in the two long-lived ${}^1 S_0 \text{ (g)}-{}^3 P_0 \text{(e)}$ clock states in a nuclear-spin polarized gas. Pauli exclusion prevents undesirable \text{e}-\text{e} inelastic collisions in the lowest band. The current experimental excited-state lifetime (limited by light scattering) is $\sim 10$ s~\cite{goban2018lifetimeLimitation}, which is 2 orders of magnitude larger than $t_c$.

The achievable SOC phase depends on $|\vec{k}_{\mathrm{C}}|=2\pi/\lambda_{\mathrm{C}}$, with $\lambda_{\mathrm{C}}$ the transition wavelength. For the ${}^1 S_0-{}^3 P_0$ states in the magic-wavelength lattice, the $\lambda_{\mathrm{C}}\approx 698$ nm clock laser naturally imparts the required SOC. To achieve the necessary $\pi$ phase in 1D, one needs to suppress tunneling along the $\hat{y}$, $\hat{z}$ lattice directions, enable tunneling along $\hat{x}$ and incline the laser so that  $a\vec{k}_{\mathrm{C}}\cdot\hat{x}=\pi$. For 2D, one enables tunneling along $\hat{x}$, $\hat{y}$, points the laser in that plane at $45^{\circ}$, and likewise inclines until the projection along both equals $\pi$. While the current magic-wavelength lattice requires a slightly larger $|\vec{k}_{\mathrm{C}}|$ for 2D, it can be adjusted through the use of accordion lattices to increase $a$, or by using a separate laser for each axis.

Alternatively, one can use two nuclear-spin states in the ${}^1 S_0$ ground-state manifold and Raman transitions to generate the desired SOC, with the one-photon detuning of the Raman lasers set sufficiently large for a long coherence time \cite{mancini2015soc}. In particular, the ${}^1 S_0-{}^1 P_1$ at $\lambda_{\mathrm{C}}\approx 461$ nm is appealing since it naturally realizes a SOC phase difference of $\approx \pi$ in each direction when the laser is oriented along the $(1,1,1)$ spatial axis, providing the framework for a 3D cluster state.

\textit{Conclusions and outlook.}
We  proposed a protocol to generate macroscopic cluster states in 3D lattice arrays of ultracold atoms via dynamical evolution. The progress of individual atom control and manipulation offered by quantum gas microscopes~\cite{bakr2009quantumGasMicroscope,parsons2015quantumGasMicroscope}, optical tweezers~\cite{endres2016tweezers} as well as the recent capability of micron-resolution spatial imaging with submillihertz frequency resolution in optical lattice clocks~\cite{campbell2017clockSeminal} are already allowing experiments to prepare high-fidelity Mott insulators needed for high quality cluster states. Combined with long-coherent times offered by AEs, our protocol can open a path for first proof-of-principle demonstrations of one-way computing schemes in the near future.

\textit{Acknowledgements.}
M.M. and A.M.R. acknowledge helpful discussions from M.A. Perlin, S.R. Muleady and P. He. M.M. acknowledges funding from the Center for Theory of Quantum Matter graduate fellowship (CTQM). R.B. acknowledges a visiting fellowship at JILA. This work is supported by the Air Force Office of Scientific Research grants FA9550-18-1-0319 and its Multidisciplinary University  Research  Initiative  grant (MURI), by the Defense Advanced Research Projects Agency (DARPA) and Army Research Office grant W911NF-16-1-0576, the National Science Foundation grant PHY-1820885, JILA-NSF grant PFC-173400, and the National Institute of Standards and Technology.

\bibliographystyle{unsrt}
\bibliography{ClusterStateBibliography.bib}

\clearpage

\onecolumngrid
\appendix

\section{Fermi-Hubbard rotated basis}
\label{app_FermiHubbardRotated}
The  Hamiltonian describing the dynamics of the long lived clock states,  $\tilde{\sigma}\in \{g,e\}$  
 trapped in the 3D optical lattice when  driven by the clock laser is given by,
\begin{equation}
    \hat{H}=-J\sum_{\langle j,k\rangle,\tilde{\sigma}}\left(\hat{b}^{\dagger}_{j,\tilde{\sigma}}\hat{b}_{k,\tilde{\sigma}}+h.c.\right)+U\sum_{j}\hat{n}_{j,g}\hat{n}_{j,e}-\frac{i \Omega}{2}\sum_{j}\left(e^{i j \phi}\hat{b}_{j,e}^{\dagger}\hat{b}_{j,g}-h.c.\right),
\end{equation}
where  $\hat{b}_{j,\tilde{\sigma}}$ creates a fermion on site $j$, and $\hat{n}_{j,\tilde{\sigma}}=\hat{b}_{j,\tilde{\sigma}}^{\dagger}\hat{b}_{j,\tilde{\sigma}}$. The phase $\phi = \phi_{j+1}-\phi_{j} = \vec{k}_{\mathrm{C}}\cdot \left(\vec{r}_{j+1}-\vec{r}_{j}\right)$ is induced by the same drive laser with wavevector $\vec{k}_{\mathrm{C}}$ acting on all sites. The value of $\phi$ describes an effective flux that characterizes the spin-orbit coupling our system exhibits.

We make a rotation to diagonalize the driving term, defining two new flavors of fermion with spin $\sigma \in \{\uparrow, \downarrow\}$,
\begin{equation}
\begin{aligned}
    \hat{c}_{j,\uparrow}&=\left(\hat{b}_{j,e}-i \hat{b}_{j,g}\right)/\sqrt{2},\\
    \hat{c}_{j,\downarrow}&=\left(\hat{b}_{j,e}+i \hat{b}_{j,g}\right)/\sqrt{2}.
    \end{aligned}
\end{equation}
In this basis, the Hamiltonian becomes,
\begin{equation}
    \hat{H}=-J \sum_{\langle j,k\rangle,\sigma}\left(\hat{c}^{\dagger}_{j,\sigma}\hat{c}_{k,\sigma}+h.c.\right)+U\sum_{j}\hat{n}_{j,\uparrow}\hat{n}_{j,\downarrow}+ \frac{\Omega}{2}\sum_{j}\left[\cos(j\phi)\left(\hat{c}_{j,\uparrow}^{\dagger}c_{j,\uparrow}-\hat{c}_{j,\downarrow}^{\dagger}\hat{c}_{j,\downarrow}\right)-i \sin(j\phi)\left(\hat{c}_{j,\uparrow}^{\dagger}\hat{c}_{j,\downarrow}-\hat{c}_{j,\downarrow}^{\dagger}\hat{c}_{j,\uparrow}\right)\right].
\end{equation}

If we now assume that the laser is aimed such that $\phi = \pi$, the second half of the driving term vanishes, and we are left with Eq.~\eqref{eq_FermiHubbard} in the main text.

\section{Superexchange}
\label{app_Superexchange}

We compute the superexchange interaction for our system. Since tunneling is nearest-neighbour, at half filling we can obtain the relevant physics from considering a two-site system ($j=1,2$). Starting with Eq.~\eqref{eq_FermiHubbard}, we make the gauge transformation $\hat{d}_{j\uparrow}=\hat{c}_{j\uparrow}$, $\hat{d}_{j\downarrow}=(-1)^{j}\hat{c}_{j\downarrow}$, which yields the two-site Hamiltonian,
\begin{equation}
\label{app_eq_FermiHubbardGauged}
    \hat{H}_{2}=J\left(\hat{d}_{1\uparrow}^{\dagger}\hat{d}_{2\downarrow}+\hat{d}_{1\downarrow}^{\dagger}\hat{d}_{2\uparrow}+h.c.\right)+U\sum_{j=1}^{2}\hat{n}_{j\uparrow}\hat{n}_{j\downarrow}+\frac{\Omega}{2}\sum_{j=1}^{2}\left(\hat{n}_{j\uparrow}-\hat{n}_{j\downarrow}\right),
\end{equation}
with $\hat{n}_{j\sigma}=\hat{d}_{j\sigma}^{\dagger}\hat{d}_{j\sigma}$ here. The $J$-proportional tunneling perturbs the diagonal $U, \Omega$ terms. We split the Fock states of the system into two manifolds,
\begin{equation}
\begin{aligned}
    \mathbbm{E}_{0}&=\{\ket{\uparrow,\uparrow}, \ket{\uparrow,\downarrow},\ket{\downarrow,\uparrow},\ket{\downarrow,\downarrow}\},\\
    \mathbbm{E}_{\mathrm{U}}&=\{\ket{\uparrow\downarrow,0},\ket{0,\uparrow\downarrow}\},
\end{aligned}
\end{equation}
where we assume a basis ordering $\ket{n_{1\uparrow},n_{1\downarrow},n_{2\uparrow},n_{2\downarrow}}$ in the Fock states (since hopping between states acquires a minus sign for every particle in-between). Assuming $U/J \gg 1$ and no initial double-occupancies, we drop $\mathbbm{E}_{\mathrm{U}}$ from the Hilbert space and define an effective Hamiltonian acting on $\mathbbm{E}_{0}$,
\begin{equation}
\label{eq_Superexchange}
\bra{i}\hat{H}_{\mathrm{se,2}}\ket{j}=\sum_{k\in\mathbbm{E}_{\mathrm{U}}}\frac{\bra{i}\hat{H}_{J}\ket{k}\bra{k}\hat{H}_{J}\ket{j}}{\frac{1}{2}(E_i+E_j)-U},
\end{equation}
where $\hat{H}_J$ is the tunneling term in Eq.~\eqref{app_eq_FermiHubbardGauged}, and the states $\ket{i}, \ket{j} \in \mathbbm{E}_{0}$ have energy $E_i$, $E_j$ (in the written basis, these are $\Omega, 0, 0, -\Omega$ respectively). The two-site superexchange Hamiltonian in the basis of $\mathbbm{E}_{0}$ yields a $4\times 4$ matrix,
\begin{equation}
    \hat{H}_{\mathrm{se,2}}=2J^2\left(\begin{array}{cccc}\frac{1}{\Omega-U}&0&0&\frac{1}{U}\\0&0&0&0\\0&0&0&0\\\frac{1}{U}&0&0&\frac{-1}{\Omega+U}\end{array}\right).
\end{equation}
We map this singly-occupied Hilbert space to that of a spin-1/2 model. Spin operators are defined using standard conventions, $\hat{S}_{j}^{z}=(\hat{n}_{j\uparrow}-\hat{n}_{j\downarrow})/2$, $\hat{S}_{j}^{+}=(\hat{S}_{j}^{-})^{\dagger}=\hat{d}_{j\uparrow}^{\dagger}\hat{d}_{j\downarrow}$. We find a nearest-neighbour spin-spin interaction,
\begin{equation}
\label{eq_appSuperexchangeTwoSites}
    \hat{H}_{\mathrm{se,2}}=\frac{4J^2 U}{\Omega^2-U^2}\hat{S}_{1}^{z}\hat{S}_{2}^{z}+\frac{2J^2}{U}\left(\hat{S}_{1}^{+}\hat{S}_{2}^{+}+h.c.\right)+\left(\Omega+\frac{2J^2 \Omega}{\Omega^2-U^2}\right)\left(\hat{S}_{1}^{z}+\hat{S}_{2}^{z}\right),
\end{equation}
Note that the bare $\Omega$ single particle term did not come from superexchange, and was instead kept from the original Hamiltonian. This interaction is present across every link of the lattice, yielding Eq~\eqref{eq_SuperexchangeHamiltonian} in the main text. The superexchange-derived single-particle term will apply twice to every site for every dimension of tunneling, yielding the additional multiplicative factor in the main text result.

\section{Spin-echo}
\label{app_SpinEcho}
We provide more details on how our spin-echo protocol replicates an Ising model in the presence of both single-particle terms and other interactions. Since all of our Hamiltonian terms are local or nearest-neighbour to good approximation, we can gain an understanding of the physics by considering a two-site system. The full superexchange Hamiltonian for two sites ($j=1,2$) is given by Eq.~\eqref{eq_appSuperexchangeTwoSites}, and we seek to isolate the $\hat{S}_{1}^{z}\hat{S}_{2}^{z}$ term.

Assuming a deep lattice and strong drive ($U/J,\> \Omega/J \gg 1$), the system's unitary time-evolution can be written as,
\begin{equation}
\begin{aligned}
    e^{-i \hat{H}_{\mathrm{se,2}}t}=&\left(
\begin{array}{cccc}
 e^{-i t \left(J_{\text{zz}}/4+\Omega\sqrt{1+J_{\text{zz}}/U} \right)} & 0 & 0 & -\frac{2 i J^2}{U \Omega} e^{-i t J_{\text{zz}}/4}\sin \left(t
   \Omega  \sqrt{1+J_{\text{zz}}/U}\right)\\
 0 & e^{i t J_{\text{zz}}/4} & 0 & 0 \\
 0 & 0 & e^{i t J_{\text{zz}}/4} & 0 \\
 -\frac{2 i J^2}{U \Omega}e^{-i t J_{\text{zz}}/4} \sin \left(t \Omega  \sqrt{1+J_{\text{zz}}/U}\right) & 0 & 0 & e^{-i t
   \left(J_{\text{zz}}/4-\Omega  \sqrt{1+J_{\text{zz}}/U}\right)} \\
\end{array}
\right)+\mathcal{O}\left(\frac{J^{4}}{U^2\Omega^2}\right),
\end{aligned}
\end{equation}
where $J_{\text{zz}}=4J^2U/(\Omega^2-U^2)$ as before.

The additional unwanted interactions $\hat{S}^{+}\hat{S}^{+}+h.c.$ are contained in the off-diagonal terms. The key observation is that because of maximal SOC flux, the unwanted interactions couple equal-spin states instead of opposite ones (in which case they would take the form $\hat{S}^{+}\hat{S}^{-}$ instead). Consequently, the nonzero off-diagonal matrix elements are between $\ket{\uparrow\uparrow},\ket{\downarrow\downarrow}$ instead of $\ket{\uparrow\downarrow},\ket{\downarrow\uparrow}$. These are the same states that pick up a high-frequency phase $\sim e^{-2 i \Omega t}$ from the laser drive. As a result, off-diagonal terms in the unitary become suppressed by $\sim J/\Omega$. Since they were already proportional to $\sim J/U$, their prefactor becomes $\sim J^2/U\Omega$, which is very small ($\sim 10^{-4}$) in our chosen parameter regime. When we compute fidelity between this model and an ideal Ising unitary, the effect of the off-diagonal terms is further squared due to quantum interference. Thus fidelity drop due to the unwanted interactions scales as $J^4/U^2\Omega^2$, which is effectively negligible assuming half filling.

The single-particle terms are only present in the diagonal $\Omega \sqrt{1+J_{zz}/U}$-proportional phase contributions, and are removed with the spin-echo. This leaves us with just the Ising model.

\section{Fermi-Hubbard to superexchange model agreement}
\label{app_ModelAgreement}
\begin{figure}[h!]
\centering
\includegraphics[width=0.65\linewidth]{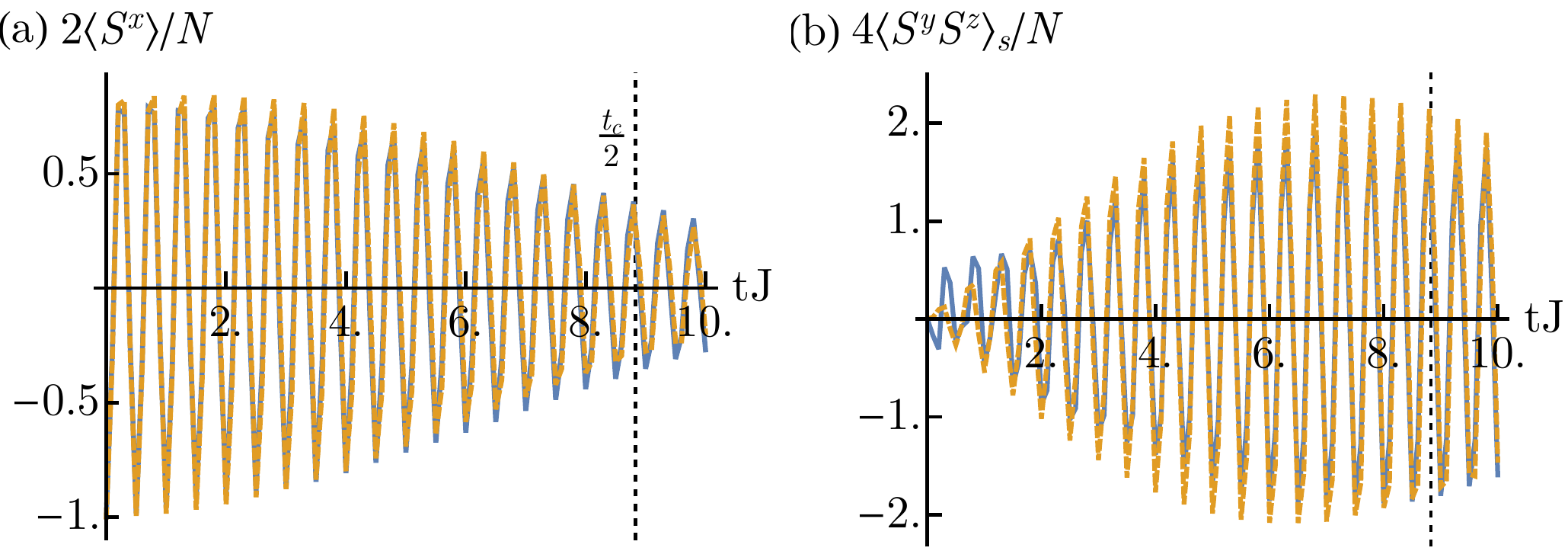}
\caption{Dynamics of collective observables $\langle \hat{S}^{x}\rangle$ and $\langle \hat{S}^{y}\hat{S}^{z}\rangle_{s} = \langle \hat{S}^{y}\hat{S}^{z}+h.c.\rangle$ for the Fermi-Hubbard model of Eq.~\eqref{eq_FermiHubbard} (blue) and the superexchange model of Eq.~\eqref{eq_SuperexchangeHamiltonian} (orange). We use a 2D lattice of dimension $L=4\times 2$, with periodic boundary conditions along the 4-site axis. Parameters are $U/J = 40$, $\Omega/J = 50$, leading to a cluster time of $t_{c}\approx 17/J$. Note that plots do not capture all oscillations due to the long timescale, but nonetheless affirm the agreement between the dynamics.}
\label{fig_Comparison}
\end{figure}

The Fermi-Hubbard model of Eq.~\eqref{eq_FermiHubbard} and the superexchange model of Eq.~\eqref{eq_SuperexchangeHamiltonian} will agree provided $U/J \gg 1$, no double-occupancies in the initial state, and $|\Omega-U|/J \gg 1$. The last condition is necessary to avoid higher-order (4th, 6th, etc.) superexchange, which will modify the interaction by adding non-Ising terms. Fig.~\ref{fig_Comparison} compares dynamics involving collective observables $\hat{S}^{\alpha}=\sum_{j}\hat{S}_{j}^{\alpha}$, finding good agreement in this parameter regime for the relevant timescales.

\section{Spin-1 superexchange}
\label{app_Spin1}
When holes are present, conventional superexchange is insufficient to describe the dynamics. While one can in principle solve the Fermi-Hubbard model directly, the resulting Hilbert space scales as $4^{L}$ near unit filling ($L$ is number of sites), which is too steep to capture the relevant physics numerically. We instead derive an effective spin-1 superexchange model, for which the Hilbert space will scale as $3^{L}$ which is more tractable. The spin-up and spin-down states of the atoms will correspond to $m_s=+1$ and $m_s=-1$ spin angular momentum eigenstates, while a hole will be the $m_s = 0$ state. We work in the basis of gauge-transformed operators $\hat{d}_{j\uparrow}=\hat{c}_{j\uparrow}$, $\hat{d}_{j\downarrow}=(-1)^{j}\hat{c}_{j\downarrow}$, and associate them as follows:
\begin{equation}
\hat{d}_{j\uparrow}^{\dagger}\ket{0}_{j} \to \ket{m_s = 1}_{j}, \>\> \ket{0}_{j} \to \ket{m_s = 0}_{j}, \>\> \hat{d}_{j\downarrow}^{\dagger}\ket{0}_{j} \to \ket{m_s = -1}_{j}.
\end{equation}

We derive a superexchange interaction using this mapping. The technique used is to once again consider all Fock states of two adjacent sites, removing double-occupancies, and apply Eq.~\eqref{eq_Superexchange}. There will be 9 states this time, yielding a $9\times 9$ matrix that can be mapped to spin-1 operator tensor products. We get the following Hamiltonian,
\begin{equation}
\begin{aligned}
\label{eq_Spin1HamiltonianDrive}
\hat{H}_{\mathrm{se,h}}&=\frac{J}{2}\sum_{\langle j,k\rangle}\left(\hat{\Sigma}_{j}^{z}\hat{\Sigma}_{j}^{+}\hat{\Sigma}_{k}^{+}\hat{\Sigma}_{k}^{z}+\hat{\Sigma}_{j}^{z}\hat{\Sigma}_{j}^{-}\hat{\Sigma}_{k}^{-}\hat{\Sigma}_{k}^{z}+h.c.\right)+\frac{J^2}{2U}\sum_{\langle j,k\rangle}\left(\hat{\Sigma}_{j}^{+}\hat{\Sigma}_{j}^{+}\hat{\Sigma}_{k}^{+}\hat{\Sigma}_{k}^{+}+h.c.\right)\\
&+\frac{J^2}{\Omega^2-U^2}\sum_{\langle j,k\rangle}\left[U\hat{\Sigma}_{j}^{z}\hat{\Sigma}_{k}^{z}+U \hat{\Sigma}_{j}^{z}\hat{\Sigma}_{j}^{z}\hat{\Sigma}_{k}^{z}\hat{\Sigma}_{k}^{z}+\Omega \left(\hat{\Sigma}_{j}^{z}\hat{\Sigma}_{j}^{z}\hat{\Sigma}_{k}^{z}+\hat{\Sigma}_{j}^{z}\hat{\Sigma}_{k}^{z}\hat{\Sigma}_{k}^{z}\right)\right],
\end{aligned}
\end{equation}
where the spin-1 operators are,
\begin{equation}
\hat{\Sigma}_{j}^{z}=\left(\begin{array}{ccc}1&0&0\\0&0&0\\0&0&-1 \end{array}\right),\>\>\hat{\Sigma}_{j}^{+}=\sqrt{2}\left(\begin{array}{ccc}0&1&0\\0&0&1\\0&0&0 \end{array}\right),\>\>\Sigma_{j}^{-}=\left(\Sigma_{j}^{+}\right)^{\dagger}.
\end{equation}

While this mapping works well in the same parameter regime as regular superexchange for 1D tunneling, higher dimensions require another correction. When a fermion tunnels, it picks up a minus sign for every atom it jumps over in the Fock basis ordering, which is not captured by the spin-1 operators. In 1D, this is not a problem, because we order our basis by sites and thus cannot jump over a particle assuming at most 1 atom per site. In 2D or 3D, however, nearest-neighbour tunneling can still pick up a sign because particles effectively jump over entire rows or columns (since the Fock ordering remains a 1D vector).

These minus signs can be expressed by modifying the four-term interactions of the direct tunneling above, i.e. the $J/2$-proportional terms. What we do is change the operators as follows,
\begin{equation}
\begin{aligned}
\frac{J}{2}\hat{\Sigma}_{j}^{z}\hat{\Sigma}_{j}^{\pm} \hat{\Sigma}_{k}^{\pm}\hat{\Sigma}_{k}^{z}&\to \frac{J}{2}\hat{\Sigma}_{j}^{z}\hat{\Sigma}_{j}^{\pm}\hat{P}(j,k)\hat{\Sigma}_{k}^{\pm}\hat{\Sigma}_{k}^{z},\\
\hat{P}(j,k)&=\exp\left[i\pi\sum_{p\in \text{mid}(j,k)}(\hat{\Sigma}_{p}^{z})^2\right],
\end{aligned}
\end{equation}
where $\text{mid}(j,k)$ is all sites between $j$ and $k$ according to the Fock basis ordering. The operator $\hat{P}$ effectively counts the number of particles between the tunneling sites, and picks up a minus sign for each one of them. This replicates the fermionic statistics. Since the required operator is diagonal, it does not create significant numerical cost.

As an example, consider a 2D tunneling model, and index our sites by (m,n) for the x and y components respectively. We further assume the Fock basis is ordered (1,1), (2,1), ... (L,1), (1,2), (2,2), ... (L,2), etc. Then tunneling along the x-direction does not pick up any additional factors. However, tunneling along the y-direction will pick one up. In particular, say we tunnel from (m,n) to (m,n+1). Then, $\text{mid}[(m,n),(m,n+1)]=\{(m+1,n),(m+2,n) \dots (L,n), (1,n+1),(2,n+1), \dots (m-1,n+1)\}$. This only needs to be applied to the direct tunneling term, as the others involve the superexchange and do not need to worry over statistics.

The one remaining thing that this model does not capture is double-jump processes, where one particle tunnels over two sites at once, or two particles tunnel simultaneously. While some of these processes may be on-resonant like our desired interaction $J_{zz}$, we empirically find that their effect should remain low provided the system's doping is low.

\section{System size and external confinement}
\label{app_SizeAndConfinement}
\begin{figure}[h!]
\centering
\includegraphics[width=0.8\linewidth]{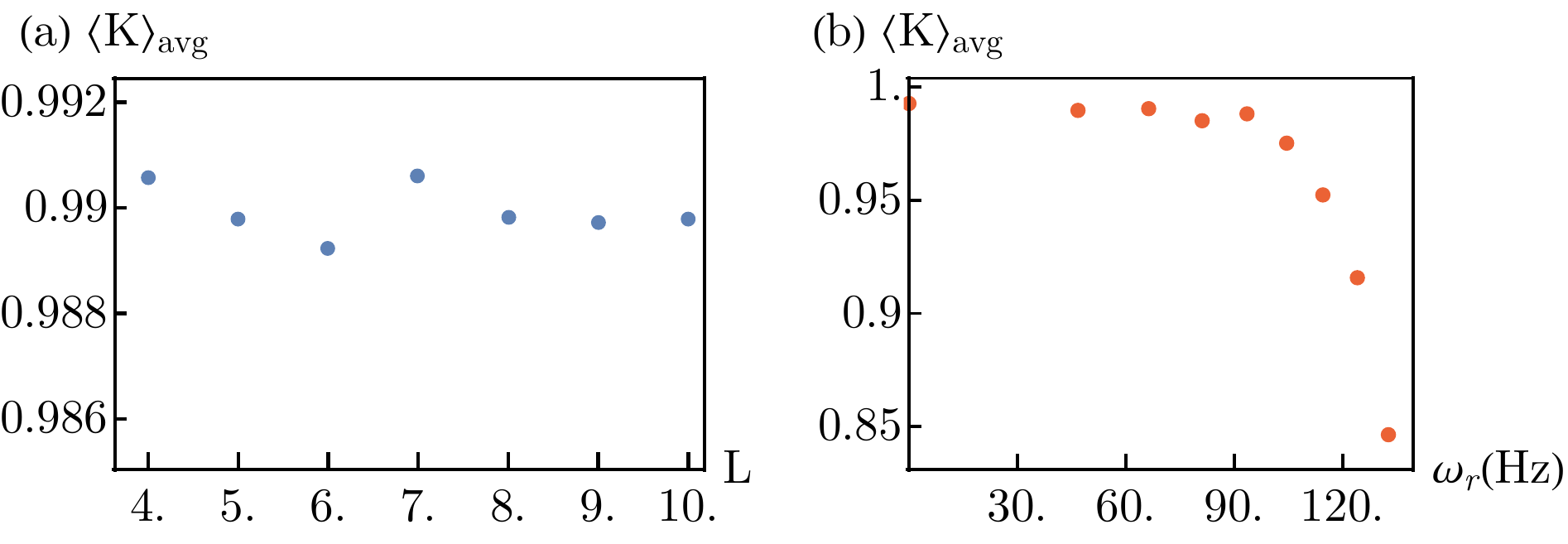}
\caption{(a) Average stabilizer values for increasing system size. Parameters are $J/(2\pi)=22$ Hz, $U/J = 115$, $\Omega/J = 140$. (b) Average stabilizer values with external trapping confinement, c.f. Eq.~\eqref{app_eq_ExternalConfinement}. System size is $L=9$, with $j_0 = 5$ in the middle. Parameters are same as in (a), along with $m = 87$ amu and $a = 406$ nm, with $\omega_r$ plotted in Hz directly.}
\label{fig_Robustness}
\end{figure}

In this supplementary, we test how our cluster stabilizers scale with system size and in the presence of external confinement. The size scaling is presented in Fig.~\ref{fig_Robustness}(a) which shows average $\langle K\rangle_{j}$ across a 1D lattice for increasing $L$. Aside from small fluctuations due to finite size effects, the value is stable in the large-$L$ limit.

The external (trapping) confinement generated by the finite beam waist of the lattice lasers can be described by an additional term to the Hamiltonian, acting as a quadratic shift in energy centered at the middle of the lattice. This additional term is, written in 1D for simplicity,
\begin{equation}
\label{app_eq_ExternalConfinement}
    \hat{H}_{\mathrm{trap}}=\frac{1}{2}m a^2 \omega_{r}^2 \sum_{j}(j-j_0)^2\left(\hat{n}_{j,\uparrow}+\hat{n}_{j,\downarrow}\right),
\end{equation}
where $j_0$ is the central lattice site, $m$ is atom mass, $a$ lattice spacing and $\omega_r$ the trapping frequency.

Fig.~\ref{fig_Robustness}(b) shows the scaling of average $\langle K \rangle_{j}$ with increasing $\omega_r$ for a 1D lattice. The number of lattice sites $L$ is chosen to be odd, with $j_0$ the central site, and open boundary conditions. We use parameters of $m = 87$ amu and $a \approx 406$ nm corresponding to Strontium-87. The trapping frequency is plotted in Hz directly for easier comparison. While a more sophisticated analysis of this perturbation would require us to further optimize over $U$ and $\Omega$, we find reasonable stability when using the same parameters as the maintext figure~\ref{fig_Holes}(a).

Note that we still include the stabilizers at the edges ($j = 1,L$) to make a more objective comparison, although their functional form changes due to the boundaries. The expression in Eq.~\eqref{eq_ClusterStabilizerDefinition} is modified by dropping the missing $2\hat{S}_{j}^{z}$ term of the site that is absent, and applying an $e^{i\pi\hat{S}^{z}/2}$ pulse to the state before computing the stabilizers at the edge. This pulse is not physical, but merely done to make the edge stabilizers match in phase to the bulk ones.

\section{Ising local measurements}
\label{app_IsingLocalMeasurements}

The result in Eq.~\eqref{eq_ClusterCollectiveProtocol} is a specific case that holds for one, two and 3D nearest-neighbour Ising models, but only at the time $t_{\mathrm{c}}$. Here we give the explicit expressions for general times in different dimensions. For 1D tunneling, a measurement of $\hat{S}_{j}^{x}$ with our $\hat{H}_{zz}$ results in the following correlations,
\begin{equation}
    e^{-i \hat{H}_{zz}t}\hat{S}_{j}^{x}e^{i \hat{H}_{zz}t}=\cos^{2}\left(\frac{J_{zz}t}{2}\right)\hat{S}_{j}^{x}-4\sin^{2}\left(\frac{J_{zz}t}{2}\right)\hat{S}_{j-1}^{z}\hat{S}_{j}^{x}\hat{S}_{j+1}^{z}+\sin\left(J_{zz}t\right)\hat{S}_{j}^{y}\left(\hat{S}_{j-1}^{z}+\hat{S}_{j+1}^{z}\right),
\end{equation}
where $j$ indexes the 1D site position. It is easy to see that for $J_{zz}t = \pi$, all but the middle term vanish, yielding the cluster correlations of interest.

For 2D tunneling, the resulting expression becomes,
\begin{equation}
\begin{aligned}
    e^{-i \hat{H}_{zz}t}\hat{S}_{j}^{x}e^{i \hat{H}_{zz}t}=&\cos^{4}\left(\frac{J_{zz}t}{2}\right)\hat{S}_{j}^{x}+16\sin^{4}\left(\frac{J_{zz}t}{2}\right)\hat{S}_{j}^{x}\prod_{k}\hat{S}_{k}^{z}+\frac{1}{4}\left[2\sin\left(J_{zz}t\right)+\sin\left(2J_{zz}t\right)\right]\hat{S}_{j}^{y}\sum_{k}\hat{S}_{k}^{z},\\
    &-\sin^{2}\left(J_{zz}t\right)\hat{S}_{j}^{x}\prod_{k<k'}\hat{S}_{k}^{z}\hat{S}_{k'}^{z}-\left[2\sin\left(J_{zz}t\right)+\sin\left(2J_{zz}t\right)\right]\hat{S}_{j}^{y}\prod_{k<k'<k''}\hat{S}_{k}^{z}\hat{S}_{k'}^{z}\hat{S}_{k''}^{z},
\end{aligned}
\end{equation}
where $k,k',k''$ products and sums are over the four nearest-neighbours of $j$ (thus for example the $k<k'$ product would include six terms). Despite the increased complexity, we see that for $J_{zz}t=\pi$, all but the second cluster correlation term still drop out. An equivalent result can be found in 3D as well.

\end{document}